\documentclass{aa}
\usepackage{txfonts}
\usepackage{graphicx}

\begin{document}

\title{Lobe Advance Velocities in the Extragalactic \\ 
       Compact Symmetric Object \object{4C\,31.04}}

\author{M. Giroletti\inst{1,2}
  \and G. Giovannini\inst{1,2}
  \and G. B. Taylor\inst{3}
  \and J. E. Conway\inst{4}
  \and L. Lara\inst{5,6}
  \and T. Venturi\inst{1}}

\offprints{M. Giroletti, \email{giroletti@ira.cnr.it}}

\institute{Istituto di Radioastronomia del CNR, via Gobetti 101, 40129
 Bologna, Italy 
 \and Dipartimento di Astronomia dell'Universit\`a di Bologna, via
 Ranzani 1, 40127 Bologna, Italy 
 \and National Radio Astronomy Observatory, P. O. Box 0, Socorro, NM 87801, USA
 \and Onsala Space Observatory, 439 92 Onsala, Sweden
 \and Dpto. Fisica Teorica y del Cosmos, Universidad de Granada, 18071 Granada, Spain
 \and Instituto de Astrofisica de Andalucia (CSIC), Apdo. 3004, 18080 Granada, Spain}

\date{Received / Accepted}

\abstract{ We report on the results of a two epoch study of the low
power Compact Symmetric Object 4C\,31.04. Observations performed with
the VLBA at 5 GHz in 1995 and 2000 have yielded images of this source
at milliarcsecond angular resolution. A central core is detected, with
bright compact hot spots and extended lobes on both
sides. Model-fitting and other analysis of the data (brightness
profile, difference map) clearly indicate that the source is
expanding. We estimate the velocity of this expansion to be (0.085
$\pm$ 0.016) mas/yr, i.e. (0.33 $\pm$ 0.06) $h_{65}^{-1} \ c$ in both
hot spots. Assuming a constant expansion velocity, we estimate the
kinematic age of the source at 550 yrs. We also study the spectral
index using VLBA observations at 1.3 GHz and MERLIN at 22 GHz. The
derived spectral age is 3000-5000 years in equipartition
conditions. The two estimates are discussed and found to be in
agreement, given present uncertainties.
\keywords{galaxies: active -- galaxies: individual (4C\,31.04) -- 
 galaxies: nuclei}}

\titlerunning{Lobe Advance Velocities in the CSO 4C\,31.04}
\authorrunning{M. Giroletti et al.}
\maketitle

\section{Introduction}

In recent years, considerable efforts have been devoted to the study
of Compact Symmetric Objects (CSOs).  The name for this class of
objects is derived from their morphology -- they have sub-galactic
dimensions but the same symmetric, two-sided aspect as classical large
scale FR\,I and FR\,II radio galaxies.  The CSOs also show other
interesting properties in the radio: (1) their spectrum can peak at
frequencies as high as a few GHz (hence their frequent classification
as Gigahertz Peaked Spectrum -- see O'Dea (\cite{ode98}), Polatidis \&
Conway (\cite{pc}) and Polatidis et al. (\cite{pol02}) for recent
reviews), (2) the rate of H\kern0.1em{\sc i}\ absorption is higher
than in common radio galaxies (e.g. Van Gorkom et al. \cite{jvg89};
Pihlstr\"om \cite{pih01}), and (3) variability and polarization are
almost non-existent (Fassnacht \& Taylor \cite{fas01}). At other
wavelengths a comprehensive study of the properties of CSO host
galaxies is still missing, although HST observations have suggested
that in at least some cases the host galaxies are not completely
relaxed (Perlman et al. \cite{per01}).

Our current understanding of CSOs suggests that they are small by
virtue of their youth (\textit{youth scenario}) and not confined by an
extremely dense medium (\textit{frustration scenario}).  If they are
young then CSOs represent the very first stage of an evolutionary
sequence (Fanti et al. \cite{fan95}; Readhead et al. \cite{rea96a},
\cite{rea96b}; Polatidis \& Conway \cite{pc}) in which a radio-quiet
galaxy turns into the host of a large, kpc-scale radio loud object. It
is therefore of great interest to (1) investigate evidence supporting
and quantifying the age of CSOs and (2) relate their current
properties to the mechanisms triggering the nuclear activity.  Indeed,
not only the triggering but also the fueling of this activity can be
studied in these objects. Since their orientation provides continuum
emission on both sides of the center of activity one can search for
either H\kern0.1em{\sc i}\ or free-free absorption (Peck et
al. \cite{pec00b}).

The radio galaxy 4C\,31.04 (B2\,0116+31) was observed with VLBI by
Wrobel \& Simon (\cite{wro86}) at 92 cm using 7 antennas.  Their image
showed two components separated by about 70 mas. Cotton et
al. (\cite{cot95}) tentatively classified this source as a low
redshift ($z=0.0592$) CSO. Giovannini et al. (2001, hereafter
\cite{gio01}) confirmed the CSO structure and identified the core with
a weak flat spectrum central component.

On the basis of spectral line VLBI observations, Conway (1996 --
hereafter \cite{con96}, \cite{con99}) has presented evidence for a
circumnuclear H\kern0.1em{\sc i}\ disk, viewed close to edge-on, whose
axis is that of the radio jet. The presence of a circumnuclear disk is
also supported by HST observations (Perlman et al. \cite{per01}),
which reveal a disk-like dusty feature of somewhat larger scale but
similar orientation to the one suggested by Conway.  The optical
nucleus is also found to have cone-like features well aligned with the
radio axis.

After a five year gap following the first observation at 5 GHz
presented in \cite{gio01}, we re-observed this source with the VLBA in
order to look for a possible expansion. After describing the
observations and data reduction in Sect.  2, we present our detailed
comparison of the two epochs and information on spectral index in
Sect.  3.  Some implications of the results are discussed in Sect.  4.
Conclusions are summarized in Sect.  5.

We assume H$_0 = 65$ km s$^{-1}$ Mpc$^{-1}$ and q$_0 = 0.5$
throughout. At the distance of 4C\,31.04 ($z=0.0592$) 1 mas
corresponds to 1.2 pc and 1 mas/yr = 3.9 $c$.
  
\section{Observations}

\subsection{5 GHz Observations}

The first epoch observations at 5 GHz were obtained on July 22, 1995
(1995.554), with the Very Long Baseline Array\footnote{The National
Radio Astronomy Observatory is operated by Associated Universities,
Inc., under cooperative agreement with the National Science
Foundation.} plus a single VLA antenna (Y1) for $\sim$5 hours.
Details about these observations are discussed in Appendix A of
\cite{gio01}.

In the second epoch we obtained $\sim$10 hrs of observing time with
the full VLBA + Y1 in full polarization mode on July 3, 2000
(2000.504). The observing frequency was centered at 4.971 GHz with 4
IFs and both polarizations recorded. Correlation was performed in
Socorro.  All calibration and fringe fitting were done within NRAO
Astronomical Image Processing System (AIPS): J0136+4751 was used as an
amplitude and leakage term calibrator, while J1751+0939, 3C395 and
J0423$-$1020 were used as calibrators for the polarization angle.

For the purposes of this work it was important to have both datasets
reduced in the same way. Thus we reanalyzed first epoch data, in order
to have the same calibration procedure for both observations. We then
exported the two datasets to DIFMAP\footnote{DIFMAP was written by
Martin Shepherd at Caltech and is part of the VLBI Caltech Software
package}, which we used to perform the editing, self calibration and
model-fitting. To properly register the two images, we used the
1995.554 image as a starting model for the self calibration of the
2000.504 data. We chose a restoring beam of 3 mas FWHM for both
epochs.

\subsection{1.3 GHz and 22 GHz Observations}

It is of great interest to compare the dynamic age -- derived from the
study of hot spot velocities, to the radiative age -- which can be
estimated from synchrotron losses.  In order to make this comparison,
observations at multiple wavelengths are needed.

Data at 1340 MHz were obtained with the VLBA + Y1 on July 7, 1995 in a
project aimed to study the nature of H\kern0.1em{\sc i}\ absorption in
CSOs. Spectral line results and a continuum map with angular
resolution of 3 mas (FWHM) were reported in \cite{con96}. Details on
data calibration and reduction will be presented in a forthcoming
paper (Conway et al., in prep.).  Here we point out only that the same
strategy used for the 5 GHz data was adopted.

We also investigated the high energy end of the spectrum by
considering a MERLIN observation at 22 GHz. Observations were made on
29th January 1998 in continuum mode with a bandwidth of 14 MHz, using
all 5 of the MERLIN telescopes available at this frequency. The
observations lasted 12 hrs. Observations were made using
phase-referencing, but only low dynamic range images were obtained
using phase-referencing so final images were made using
self-calibration. Good fits to the phase and amplitude data were
obtained. Final restoring beam was 10 mas (FWHM).

We used the 1.3 GHz data and the second epoch 5 GHz observation to map
the spectral index.  For a proper comparison we tried to achieve a
similar $(u,v)$ coverage at the two observing frequencies by cutting
the shortest baselines in the 20 cm data and the longest ones in the 6
cm data. Final images were then produced with the same angular
resolution and cell--size. The source position in the two images was
registered using the isolated, unresolved core.

The resulting $(u,v)$ coverages at 5 and 22 GHz are quite different,
so it was not practical to produce a spectral index image between
these frequencies. However, comparable short baselines are present, so
that we do not expect missing flux in either image. This allowed us to
derive an integrated spectrum for different regions of the source.

\section{Results}

\subsection{Morphology}

\begin{figure}
  \resizebox{\hsize}{!}{\includegraphics{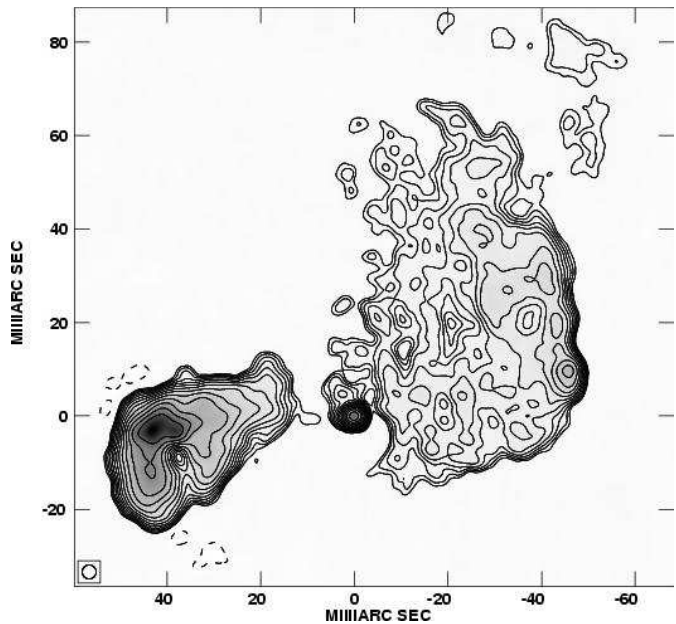}}
  \caption{The July 2000 VLBA image of 4C\,31.04 at 5 GHz.  Contours
  are drawn at $-$0.4, 0.4, 0.57, 0.8, ..., 36.2 mJy/beam by
  $\sqrt{2}$ intervals with negative contours shown dashed.  The peak
  flux density is 40.2 mJy/beam.  The grey scale range is from $-0.75$
  to 40.2 mJy/beam.  Note the compact core, symmetrically placed hot
  spots, and the unusual `hole' in the eastern lobe.}
  \label{fig1}
\end{figure}

\begin{figure}
  \resizebox{\hsize}{!}{\includegraphics{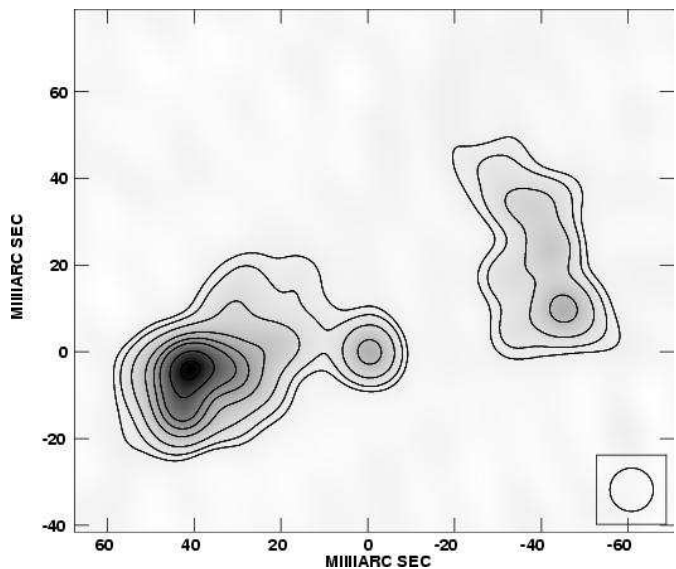}}
  \caption{MERLIN image at 22 GHz restored with 10 mas FWHM. Contours
  are drawn at $-$4, 4, 8, 16 32, 48, 64, 96, 128 mJy/beam. The peak
  flux density is 138.5 mJy/beam.  The grey scale range is from $-4.3$
  to 138.5 mJy/beam. } 
  \label{fig2}
\end{figure}

In Figure \ref{fig1} we show the final image of 4C\,31.04 at 5 GHz for
the July 2000 epoch, convolved with a circular restoring beam of 3 mas
(FWHM).  The overall structure of the source is the same as was
observed in 1995 (both at 5 and 1.3 GHz), consisting of a compact core
component with hot spots and lobes on either side. The lower
resolution (10 mas FWHM) 22 GHz MERLIN image shown in Fig. \ref{fig2}
shows the same features, except that the lobes look much fainter and
are barely detected.

At all frequencies the total source extent is $\sim$100 mas, which
corresponds to $\sim$120 pc. Total flux density in the 5 GHz maps is
$\sim 1.25$ Jy, i.e. about 90\% of values measured with single dish
telescopes (Becker et al. \cite{bec91}; Gregory \& Condon
\cite{gre91}). This yields a total power of $1.14 \times 10^{25}$ W/Hz
at 5 GHz, which makes 4C\,31.04 one of the weakest known CSOs
(Polatidis et al. \cite{pol02}).

Within Difmap, we model-fit the visibility data of the two 5 GHz
observations with a set of 10 elliptical Gaussian components. We used
large, extended components in order to pick up the lobes' flux density
as well as compact components to describe the core and hot spots.  We
fixed the parameters of the largest components (size $>$ 10 mas) and
let the most compact ones vary; results of the fits are listed in
Table \ref{tab1}. Uncertainties for each component are computed from
the ratio between its FWHM and SNR and are all within 0.1 mas.  The
most compact components, except for the core, are all located in the
hot spots regions. The western hot spot is labeled W1, while the best
fit to the eastern end is obtained using three different components --
E1, E2 (also the image peak of 40.2 mJy/beam) and E3. Given their high
brightness, position close to the edge of the lobe and compactness, we
will consider W1 and E2 as the two primary hot spots of this
source. It is less likely that either E1 or E3 -- both quite fainter
than E2 -- may be a primary hot spot; it is however possible that they
are secondary hot spots similar to those found in many extended FR\,II
sources. The innermost E3 could also be related to the presence of a
knot component in an underlying jet.

Indeed, even if a classical jet of plasma is not detected, hints of a
jet-like feature are clearly present in the enhanced brightness
visible all along the line connecting the core, and in the east hot
spot which has an elongated structure on the west side in the full
resolution image published in \cite{gio01}.  Giovaninni et al. suggest
that the source is oriented close to the plane of sky ($\theta >
75^{\circ}$), with the west lobe the one nearest to us, in agreement
with \cite{con96}.

With the increased image fidelity obtained in this second epoch we can
also confirm the presence of a `hole' in the east lobe.  This large
region of lesser brightness lies close to the east hot spot, in a
direction perpendicular to the jet. While visible in the first
observation, it looks even dimmer and more defined in this second
epoch.  In contrast, the west lobe does not present very clear
features. Being very extended and of low brightness, it cannot be
mapped with high fidelity by VLBI technique; however, we are confident
that this does not affect the compact hot spot region, which is well
detected on most baselines. Fig. \ref{fig2} and the 1.3 GHz data
(\cite{con96}) also confirm that it is indeed a compact, flat-spectrum
component.

At their outermost extents, both lobes have a very sharp edge, running
along a very wide front ($\sim 20$ mas in the eastern lobe but even
$\sim 40$ mas in the western one). The sharpness of this border is
again in agreement with this region being the place where the plasma
is advancing into the external medium as the source grows.

\subsection{Component Motions and Kinematic Age}

\begin{table*}
  \caption[]{Gaussian model and relative proper motions for
  4C\,31.04.}
\begin{center}
\begin{tabular}{l r r r r r r r r r r}
\hline
\hline
Component & Epoch & \multicolumn{1}{c}{$S$} & \multicolumn{1}{c}{$r$} & \multicolumn{1}{c}{$\theta$} &\multicolumn{1}{c}{$a$} &\multicolumn{1}{c}{$b/a$}& \multicolumn{1}{c}{$\Phi$} & \multicolumn{1}{c}{$\Delta r$} & \multicolumn{1}{c}{$v$} & p.a. \\
       &  &  (Jy) &    (mas)   &  \multicolumn{1}{c}{($^{\circ}$)}&     (mas)  &   & \multicolumn{1}{c}{($^{\circ}$)}  & (mas) & \multicolumn{1}{c}{($h^{-1}_{65}$ c)} & ($^{\circ}$)  \\
\hline
\noalign{\vskip2pt}
C\ldots  & 1995.554 & 0.020 &  0.0     &      0.0  &  0.81 &  1.00 &     8.0  \\
         & 2000.504 & 0.024 &  0.0     &      0.0  &  0.81 &  1.00 &     8.0 &  reference \\
\hline
E1\ldots & 1995.554 & 0.013 &  46.74   &     93.7  &  1.60 &  1.00 &    32.5 \\
         & 2000.504 & 0.013 &  46.82   &     93.9  &  1.60 &  1.00 &    32.5 & 0.18 $\pm$ 0.01 & 0.14 $\pm$ 0.01 & 157.6 \\
E2\ldots & 1995.554 & 0.137 &  42.52   &     93.6  &  7.20 &  0.60 & $-$28.3 \\
         & 2000.504 & 0.135 &  42.82   &     93.5  &  7.20 &  0.60 & $-$28.3 & 0.31 $\pm$ 0.04 & 0.25 $\pm$ 0.04 & 79.6 \\
E3\ldots & 1995.554 & 0.038 &  37.30   &     94.6  &  2.95 &  1.00 &  $-$6.0 \\
         & 2000.504 & 0.040 &  38.01   &     94.9  &  2.95 &  1.00 &  $-$6.0 & 0.74 $\pm$ 0.02 & 0.58 $\pm$ 0.02 & 110.2 \\
W1\ldots & 1995.554 & 0.030 &  46.08   &  $-$78.3  &  5.54 &  0.53 &     7.0 \\
         & 2000.504 & 0.027 &  46.62   &  $-$78.3  &  5.54 &  0.53 &     7.0 & 0.54 $\pm$ 0.07 & 0.43 $\pm$ 0.07 & -78.2 \\
\hline
E4\ldots & 1995.554 & 0.213 &  45.67   &    106.0  & 11.53 &  0.69 &    11.7 \\
	 & 2000.504 & 0.216 &  45.67   &    106.0  & 11.53 &  0.69 &    11.7 & fixed \\
E5\ldots & 1995.554 & 0.074 &  32.59   &    105.4  & 10.20 &  0.58 &  $-$0.9 \\
         & 2000.504 & 0.071 &  32.59   &    105.4  & 10.20 &  0.58 &  $-$0.9 & fixed \\
E6\ldots & 1995.554 & 0.209 &  33.89   &     89.4  & 20.32 &  0.40 & $-$83.2 \\
	 & 2000.504 & 0.229 &  33.89   &     89.4  & 20.32 &  0.40 & $-$83.2 & fixed \\
W2\ldots & 1995.554 & 0.333 &  45.07   &  $-$50.1  & 41.63 &  0.42 &    22.8 \\
	 & 2000.504 & 0.323 &  45.07   &  $-$50.1  & 41.63 &  0.42 &    22.8 & fixed \\
W3\ldots & 1995.554 & 0.143 &  21.91   &  $-$77.0  & 35.69 &  0.63 &    86.2 \\
         & 2000.504 & 0.162 &  21.91   &  $-$77.0  & 35.69 &  0.63 &    86.2 & fixed \\
\hline

\end{tabular}
\label{tab1}
\end{center}

NOTE -- Parameters of each Gaussian component of the model brightness
distribution: $S$, flux density; $r, \, \theta$, polar coordinates of
the center of the component relative to an arbitrary origin, with
polar angle measured from north through east; $a, \, b$, major and
minor axes of the FWHM contour; $\Phi$, position angle of the major
axis measured from north through east; $\Delta r$, relative proper
motion of the component; $v$, relative projected velocity in units of
$h^{-1}_{65}\, c$ ($h_{65}$ = H$_0$/65 km s$^{-1}$ Mpc$^{-1}$), along
the given position angle (p.a.).

\end{table*}

The first hot spot velocities in CSOs by Owsianik \& Conway
(\cite{ows98}) required time baselines of $\sim 15$ years, with VLBI
observations at 5 GHz.  By going to higher frequencies (15 and 43
GHz), Taylor et al. (\cite{tay00}) reduced this to $\sim 5$ years for
a few strong CSOs.  Currently 10 sources have been measured (Polatidis
et al. \cite{pol02}). Typical velocities are 0.2$h^{-1}_{100} \, c$
(i.e. 0.31$h^{-1}_{65} \, c$), which due to the large distance of the
sources, translates into angular motions of typically only a small
fraction of mas/yr ($\sim$32 $\mu$as/yr at $z=0.1$), thus requiring
long time intervals and high angular resolutions. We consider
observations at 5 GHz separated by 5 years.  More observations in the
future are planned to extend the time baseline and to better constrain
the results.

\begin{figure}
  \resizebox{\hsize}{!}{\includegraphics{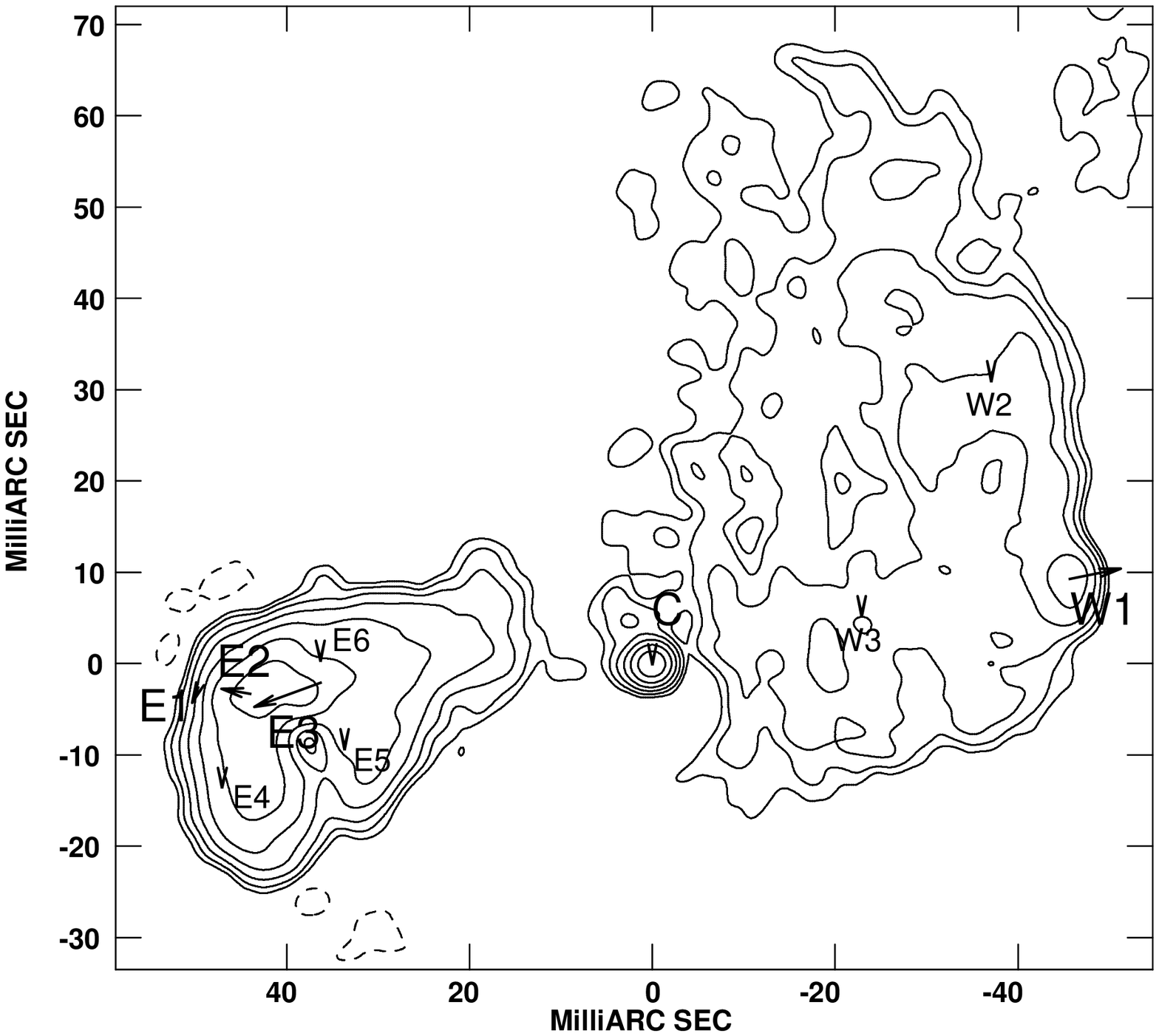}}
  \caption{The 2000 VLBA image of 4C\,31.04 at 5 GHz with arrows
  indicating the positions and motions of components derived from
  model-fitting.  The model-fit results are shown in Table 1.
  Components larger than 10 mas have been held fixed between the two
  epochs for reasons discussed in the text.  Component motions have
  been magnified by a factor of 5 for illustration purposes.  Those
  arrows pointing due south indicate that the component was held
  fixed.}
\label{fig3}
\end{figure}

\begin{figure}
  \centering
  \resizebox{!}{9cm}{\includegraphics{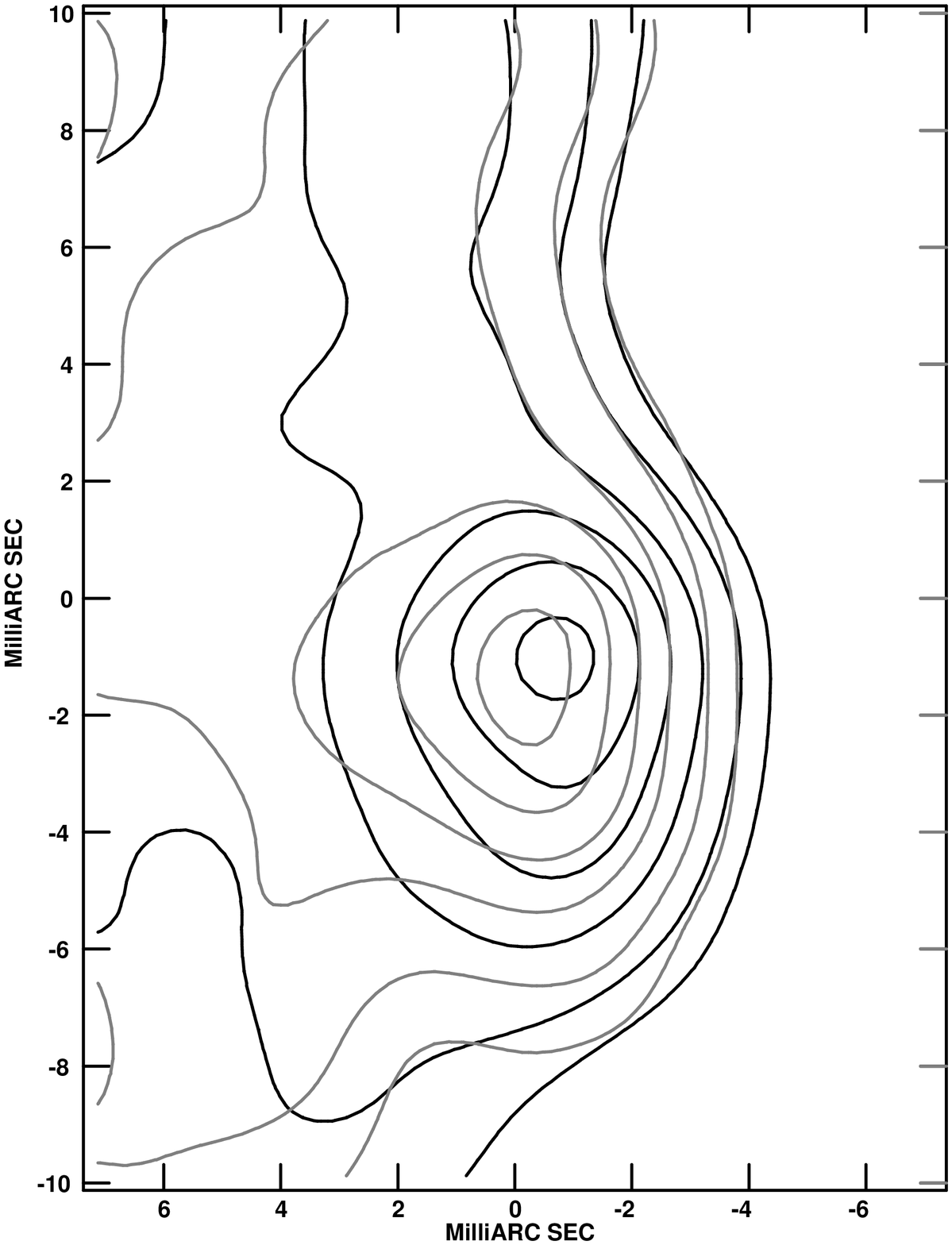}}
  \caption{Overlay of the hot spot region in the west lobe
  (W1). Contours are drawn in grey (epoch 1995.554) and black (epoch
  2000.504) at (1, 2, 4, 6, 8, 10) mJy/beam. The two images have been
  aligned on the position of the east hot spot E2.}
  \label{fig4}
\end{figure}

To measure the advance velocity in 4C\,31.04, we compare the model-fit
of the dataset in our two epochs. Fig. \ref{fig3} illustrates the
motion of all the compact components with arrows of length
proportional to the magnitude of their motion. It is apparent that all
the components have moved away from the core. In particular, the two
hot spots W1 and E2 show motions almost in the same direction as their
position angles. Furthermore, the increase in hot spots separation
becomes even more clearly visible when E2 and W1 are directly
compared. In Fig. \ref{fig4} we show a magnified inset around the west
hot spot region with contours drawn for both epochs. In this image the
2000.504 profile has been shifted by 0.3 mas westward, so that the
position of component E2 is the same in both maps (the displacement
needed in declination is smaller than the measurement errors). The
motion of the peak of W1 is clear, confirming the results of
model-fitting.

\begin{figure}
  \resizebox{\hsize}{!}{\includegraphics{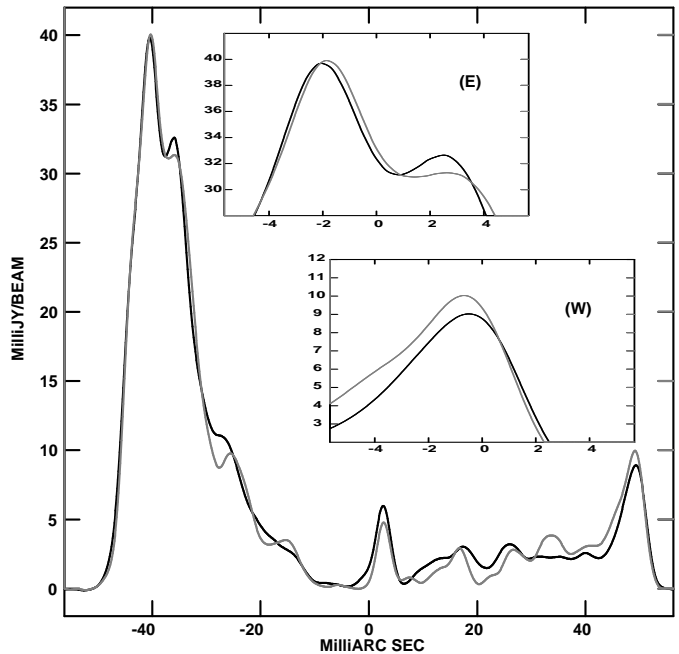}}
  \caption{A slice through the hot spots of 4C\,31.04.  The central small
  peak results from the northern edge of the core (the hot spots are 
  not perfectly symmetric about the core).  The black line
  shows the profile of the 2000.504 image, while the grey line represents 
  the profile of the 1995.554 image.  Shown inset are magnified 
  versions of the profile at the east (E) and west (W) hot spots.  }
  \label{fig5}
\end{figure}

In Fig. \ref{fig5} we show a brightness profile made in AIPS with the
task \texttt{SLICE}. The two profiles are derived from the two epochs
along the same slice. This slice cuts the two hot spots at a position
angle of $-82.6^{\circ}$ (measured north to east), passing close to
the core (but not across the core itself because of a small bend in
the jet). In the global profile one can easily see the two lobes on
the opposite sides (the east one being the brightest) and the core
region in the middle (see Sect.  \ref{sec:variability} about core
variability). The hot spot regions are magnified in the two insets,
showing that both in the eastern and the western hot spot the 2000.504
profile is shifted outward. This displacement, though small, is clear
both in the rising and in the falling part of the profile. We remind
the reader that the region between the core and the west hot spot is
poorly imaged, so that the features present in the slices in this
region are not completely reliable. It is also possible that some of
the changes are due to turbulence present in this back-flow region.

\begin{figure}
  \resizebox{\hsize}{!}{\includegraphics{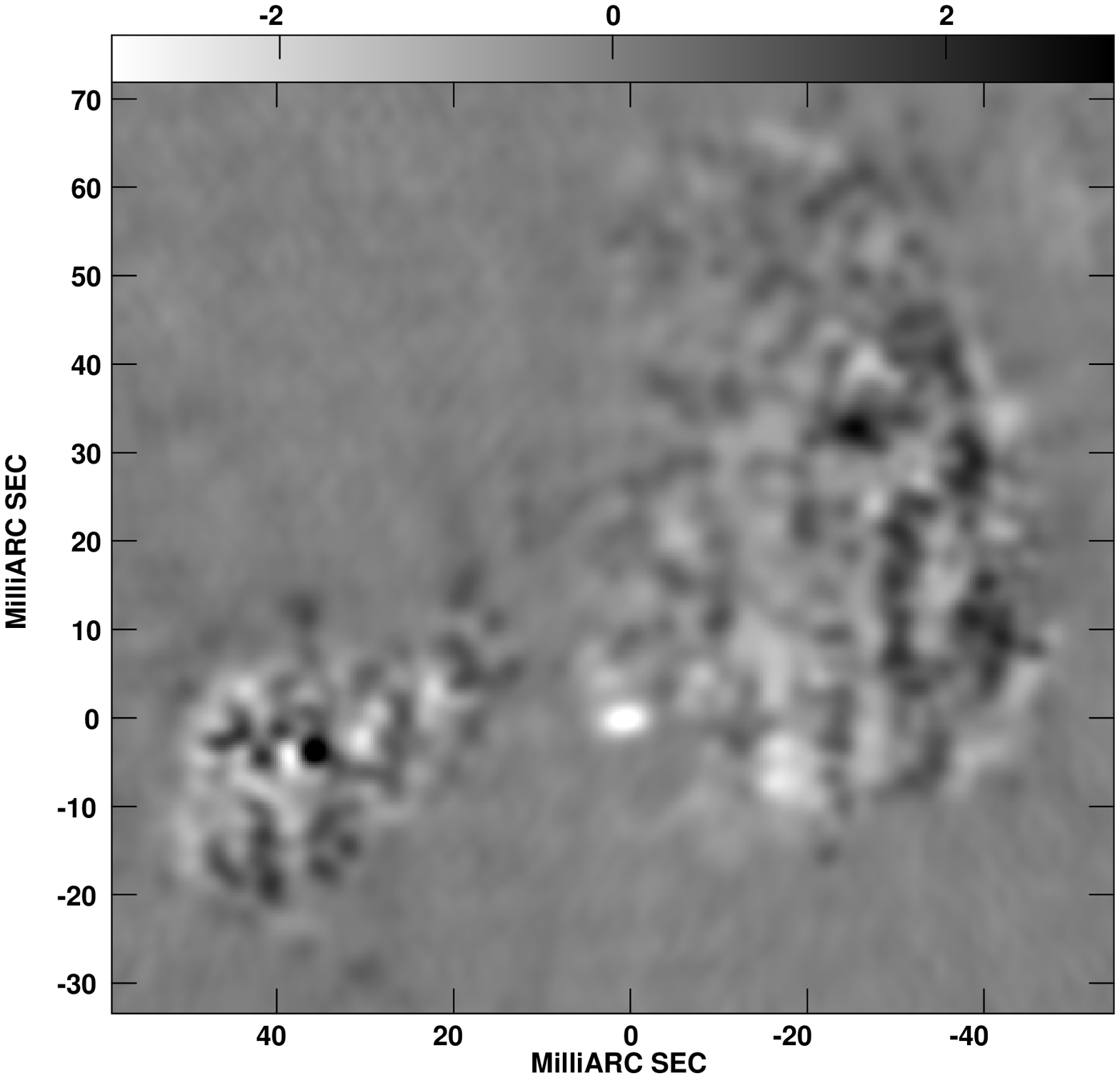}}
  \caption{The difference image computed as $S_{2000.504} - S_{1995.554}$.
  Note the change in the core flux density of $\sim$4 mJy.  One can also
  see the black+white pattern of motion in the eastern hot spot, and a 
  faint ring of white along the edge of the eastern lobe.  The ``mottled''
  pattern in the lobes is a result of the low fidelity of the image 
  in the presence of extended emission.  }
  \label{fig6}
\end{figure}

The increase in the hot spot separation implies an expansion of the
whole source. Since the lobes are too extended and faint to measure
their expansion by model-fitting, we tried to directly compare the two
images. The difference map shown in Fig. \ref{fig6} was done with the
AIPS task \texttt{COMB}. The features that appear in this image, in
particular in the eastern lobe, are clearly related to the expansion
of the source. The two close white and black spots are due to the
outward shift of the peak of component E3. Moreover, the presence of a
white stripe all around the outer edge of the lobe, though faint, is
suggestive of an expansion of the whole eastern lobe in this
direction. Note that a similar white stripe, even if less evident, is
present all around the edge of the west lobe.

As these last results (shown by Fig. \ref{fig4}, \ref{fig5},
\ref{fig6}) are drawn from the images and not from the model, we feel
that they lend support to the detection of motion and to the estimate
of its velocity derived from the model-fits.  With just two epochs
available, and a moderately short time baseline, it is difficult to
better quantify the accuracy of our measurements of component
motions. We hope to address this difficulty with future observations.

Model-fitting of the present data (see Table \ref{tab1}) provides the
most reliable information on the velocity. The east and west hot spots
present quite different displacements, $\Delta$W1 = (0.54 $\pm$ 0.07)
mas and $\Delta$E2 = (0.31 $\pm$ 0.04) mas. It is possible that such
differences may be due to a velocity asymmetry related to an
inhomogeneous ISM, but it is not possible to discuss it more deeply
with only a two epoch measurement. Since the two arms are about the
same length and the source is near to the plane of sky, we will assume
that the average velocity over the long term is the same in both
directions.

We therefore try to estimate the advance speed by averaging up the
results for the two hot spots W1 and E2.  This yields an average
motion of (0.42 $\pm$ 0.08) mas in 5 years. This indicates that each
hot spot is advancing at (0.085 $\pm$ 0.016) mas/yr, i.e. (0.33 $\pm$
0.06) $h_{65}^{-1} \, c$. The most distant component (W1) is located
46.62 mas from the core, so it would have taken 548 $\pm$ 100 years to
cover this distance at such speed. Therefore, based on kinematics, we
can estimate the source age at $\sim$ 550 years.

\subsection{Spectral Index Distribution and Spectral Age}

From the measured energy distribution of a population of electrons it
is possible -- under a few basic assumptions -- to determine how long
these electrons have been radiating. This has proved to be a useful
tool in estimating ages of compact radio sources (e.g. Murgia et
al. \cite{mur99}; Readhead et al. \cite{rea96a}). In order to perform
such a study, one needs to know the spectral index distribution and
the magnetic field in the source.

\begin{figure}
  \resizebox{\hsize}{!}{\includegraphics{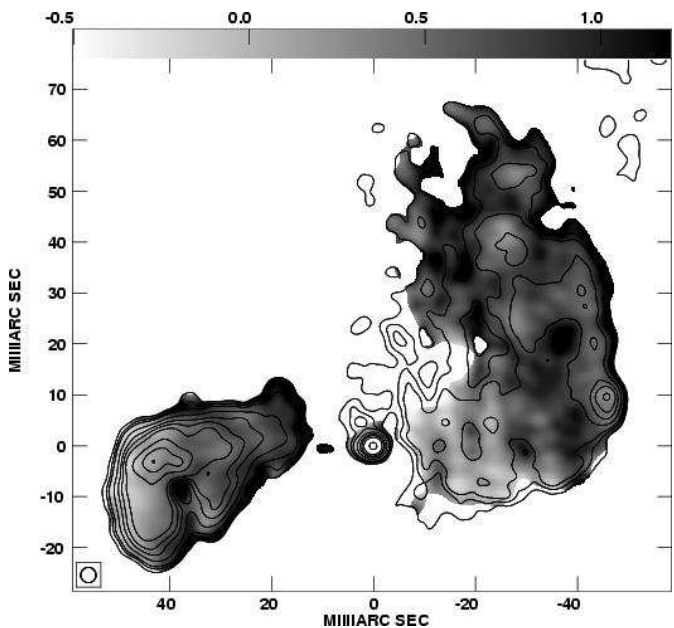}}
  \caption{The spectral index map of 4C\,31.04 at 3 mas resolution
  (FWHM). The grey scale range is $-$0.5 to 1.2. Overlaid contours from
  4.8 GHz are drawn at 0.5, 1, 1.5, 3, 5, 7, 10, 20, 30, 40 mJy/beam.}
  \label{fig7}
\end{figure}

The image in Fig. \ref{fig7} shows the spectral index for 4C\,31.04
between 1.3 and 4.8 GHz at a resolution of 3.0 mas (FWHM). The core
spectrum is inverted\footnote{Spectral index is defined according to
$S(\nu) \propto \nu^{- \alpha}$.} ($ \alpha = - 0.6 \pm 0.05$). This
is probably due to self-absorption, even after accounting for possible
flux density variability (see Sect.  \ref{sec:variability}). In the
east lobe, the hot spot region shows a flat spectrum ($\alpha$ $\sim$
0.1 $\pm$ 0.01) with a constant and uniform steepening toward the core
direction, up to $\alpha$ $\sim$ 1.0 ($\pm$ 0.02) at 20 mas from the
core, just before the `gap' between the core and lobe. The spectral
index distribution is more complicated in the west lobe perhaps on
account of a stronger interaction of the radio plasma with the ISM, as
suggested also by the lobe morphology (see Sect.
\ref{sec:morphology}). The region of the faint hot spot is still flat
($\alpha \sim 0.25 \pm 0.01$) but the spectral index steepening is
more irregular moving toward the core and to the north.  The steepest
values in the more external northern regions are as high as
1.0-1.2. In the core-hot spot direction the region from 10 to 30 mas
shows again a flat spectrum (0.1-0.4) suggesting a local interaction
between radio plasma and the dense ISM. Imaging problems in this low
brightness region could also play a role.

In this high resolution image, one can see the region of the hole
close to the east hot spot. The spectral index is very steep here, not
only in the deep hole itself -- where it is as steep as 1.7 $\pm$ 0.04
-- but also in external lobe region, where we find $\alpha$ = 1.0
$\pm$ 0.04.

In a population of electrons with an initial power-law energy
distribution, synchrotron losses produce a curvature in the original
straight power-law radio spectrum. The curvature manifests itself at a
critical frequency related to the electron age. The shape of the
spectrum and the critical frequency crucially depend on the evolution
of the electron pitch angle distribution with time (Pacholczyk
\cite{pac70}).  We used the Jaffe-Perola model which assumes a
redistribution of electron pitch angles on time scales short compared
with their radiative lifetimes (Jaffe \& Perola, \cite{jaf73}).  To
derive the critical frequency from the spectral index information, we
used the Synage program (Murgia \& Fanti \cite{mur96}).  Since a
spectral age measurement can give only the time since electrons were
accelerated, we avoided the core and hot spot regions where local
turbulence and electron re-acceleration is likely to be present, and
investigated the steep-spectrum, relaxed regions where the `oldest'
electrons are present. The break frequency was found to be $\sim$7.4
GHz in the east lobe at $\sim$20 mas from the core, and between 6.1
and 7.2 GHz in the northern region of the west lobe.

\begin{figure}
  \resizebox{\hsize}{!}{\includegraphics{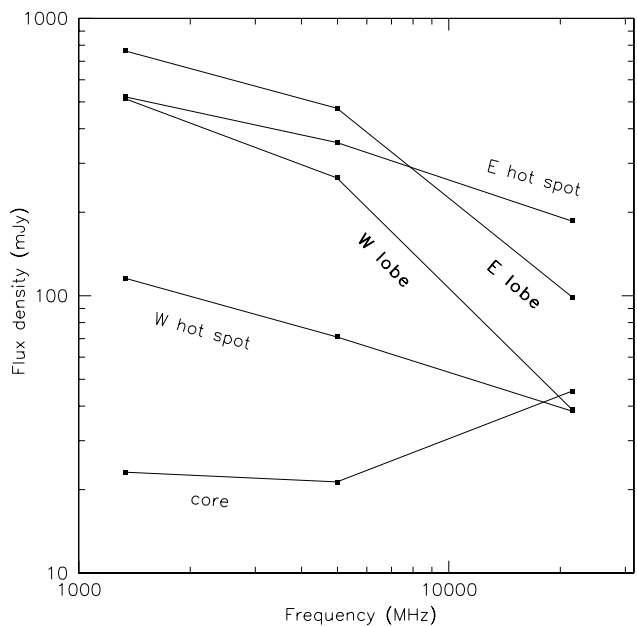}}
  \caption{The spectrum of lobes, hot spots and core. The steepening
  of both lobes is clearly visible above 5 GHz.}
  \label{fig8}
\end{figure}

The MERLIN image in Fig. \ref{fig2} lends strong support to this
estimate, showing that very little of the lobe emission is detected at
22 GHz. This cannot be due to a lack of short spacings since the
MERLIN has baselines down to 6 km. It follows that the lobes are
undetectable because of spectral steepening. Fig. \ref{fig8} shows the
spectrum for the two hot spots, the lobes and the core. The steepening
between 5 and 22 GHz in the lobes supports the presence of a spectral
break around 5-7 GHz.

The local magnetic field, which is also needed to derive the lifetime
of radiating electrons, was calculated assuming equipartition. We note
that a recent X-ray study of a relatively compact FR\,II radio source
(Brunetti et al. \cite{bru01}) indicates that the magnetic field in
the radio lobes is very near equipartition; equipartition conditions
in GPS are also preferred by Snellen \& Schilizzi (\cite{sne02}).
Standard formulae (Pacholczyk \cite{pac70}) were used under the
assumption that relativistic particles and magnetic fields occupy the
same volume ($\phi = 1$), and that the amount of energy in heavy
particles is the same as that in electrons ($k = 1$). We assumed
cutoffs in the relativistic particle energies at 10 MHz and 100
GHz. The major uncertainty in the estimation of the equipartition
field strength comes from the assumption of the source depth which we
assumed to be equal to the smaller transverse dimension of the lobe
regions where we estimated the magnetic field.

With these assumptions the derived equipartition magnetic field is 4.1
milligauss in the east hot spot region, decreasing to 3.3 milligauss
in the direction of the core. In the less bright west hot spot we find
3.5 milligauss and even smaller values (2.2-2.7 milligauss) in the
extended W lobe regions.  The uncertainties are mainly due to the
non-uniform spectrum present in this region.  With these results the
estimated radiative age of older emitting electrons is 3.0 $\times$
10$^3$ yrs in the east lobe (near the core) and 4.5-4.9 $\times$
10$^3$ yrs in the northern region of the west lobe. An estimate of the
uncertainty of these values is not obvious since errors are mainly
model dependent.

\subsection{Polarization}

After removing the instrumental polarization we did not detect any
polarized signal from the source.  The $3\sigma$ upper limit on the
polarized flux density is $<0.140$ mJy/beam at 5 GHz.  This limit
corresponds to $< 0.3$\% polarization at the brighter, eastern hot
spot, to $< 0.6$\% polarization at the core, and $< 3$\% at typical
locations in the lobes.

\subsection{Variability}
\label{sec:variability}
The sum of model-fit components in the two epochs is 1.21 and 1.24 Jy
respectively, which gives an insignificant difference of just 2\%.
Given that many of the components are extended, such a small variation
in the total flux density over 5 years seems reasonable.  The core
component is compact (size $\sim 0.81$ mas) and could be more
variable.  In fact from the model-fit results (Table \ref{tab1}) the
core varies from 20 $\pm$ 1 mJy in 1995 to 24 $\pm$ 1 mJy in 2000.
Although this variation is only significant at the 3$\sigma$ level, if
it is real then it indicates a change by 20\%.  This variation is
consistent with the limits on core variability implied by the
integrated intensity observations of Fassnacht \& Taylor
(\cite{fas01}) from a small sample of CSOs studied over the course of
eight months.  Fassnacht \& Taylor found the rms variation in the
integrated flux density to be 0.7\%.  Assuming all that variation was
due to the core component, and given an average core fraction of 0.03
(Taylor et al \cite{tay96}), leads to an estimate for the core
variability in CSOs of less than 23\%.  Better limits on the
variability of core components in CSOs will require VLBI monitoring
campaigns, rather than low resolution studies only sensitive to the
integrated flux density.  The radio galaxy 4C\,31.04 is an excellent
candidate for such a study since the core is comparatively strong and
isolated from nearby jet components.

\section{Discussion}

\subsection{Ages of Compact Symmetric Objects}
\label{sec:ages}
We derive the kinematic age of 4C\,31.04 to be 550 years.  This
estimate is within the 300 to 1200 year range of ages found in the
literature (Polatidis \& Conway \cite{pol02}; Owsianik \& Conway
\cite{ows98}; Taylor et al. \cite{tay00}).  It is interesting to
ponder the narrowness of this range.  Even younger sources would need
to be smaller, and/or have faster hot spot advance speeds.  Currently
there is a lower limit on the size of a source of a few mas ($\sim 10$
pc) that we can readily classify as a CSO by identifying a core
component between two hot spots.  Assuming that the hot spot advance
speed through the AGN environment is similar ($0.2 \, c$) for an
ultra-compact CSO (the denser environment on such small scales should
make this a conservative assumption) then we can derive a lower limit
to the age of observed CSOs of 100 years.  There is also an
observational upper limit on the size of CSOs.  This is the size
beyond which the CSO becomes difficult to observe with VLBI arrays
since an interferometer is only sensitive to certain spatial
frequencies.  VLBI arrays have difficulty imaging components larger
than about 40 beams, or sources lacking a strong compact component.
By the time CSOs reach 1 kpc in size the lobes will be largely
resolved out and the hot spots will be resolved to the extent that a
direct detection is impossible without phase referencing.  Velocity
measurements of heavily resolved components are also much more
difficult.  These parameters give a conservative upper limit on the
source age of 10000 years.  Future advances in Space VLBI may discover
radio sources younger than 100 years, and the E-MERLIN and EVLA
projects may provide sufficient resolution and sensitivity to extend
the upper boundary of sources for which we can determine a kinematic
age. We also note that the advance speed of both east and west hot
spots in 4C\,31.04 at 0.33 $h^{-1}_{65} \, c$ are at the high end of
those measured in CSOs to date, corresponding to a hot spot separation
velocity of 0.39 $h^{-1}_{100} \, c$. This should be confirmed with
future epochs.

It is of interest to compare the dynamic age of 550 years -- derived
from the study of hot spot velocities, to the radiative age -- which
we estimate from synchrotron losses in 4C\,31.04 to be $\sim4000$
years.  The radiative age is nearly an order of magnitude larger than
the dynamic age estimate, indicating that some of our assumptions are
in error.

First, we note that the kinematic age estimate rests on the assumption
that the velocity has been constant over the lifetime of the
source. Although Readhead et al (\cite{rea96b}) argue that only mild
evolution in velocity is likely, this assumption may not hold in
4C\,31.04. As discussed in Sect.  \ref{sec:morphology}, there is
evidence from the radio morphology of a strong interaction between the
jet and the ISM. This could be making the jet wander around within the
lobe, forming new hot spots which drill quickly out for some brief
time before being shut off. Further suggestion that the kinematic age
may be an underestimate comes from the west lobe dimension: the
weakest detectable emission northward is about 60 mas which would
require a lobe expanding at $0.41 \, c$.

On the other hand, it is possible that the radiative age is
overestimated by the equipartition field strength or the value of the
critical frequency. While the 22 GHz data suggest that the value of
the critical frequency is quite accurate, we note that the
equipartition field could easily be higher if either $k>1$ or $\phi <
1$. In this case the radiative age would be much more similar to the
kinematic estimate; for instance $H_\mathrm{eq} \sim 6$ milligauss
(given by $k=20$ and $\phi = 0.7$) would yield $\sim 1000$
years. Furthermore, this result was obtained assuming no energy losses
due to Inverse Compton interactions. Even if scattering with the
microwave background can be neglected with no harm, it is possible
that in such a small source some interaction is present between
relativistic electrons in the lobes and photons from the active
core. The radiative age would then be smaller than calculated.

It is not easy to infer from the present data which estimate is the
most reliable. We hope to address the question of the dynamic age with
additional observations in the next years. Other studies of magnetic
fields in CSOs would help to understand how correct are the
assumptions going into the radiative age estimate. Also, high
resolution low frequency data could help to estimate $H$ from
self-absorption analysis, while high frequency data will allow a
better estimate of the critical frequency.

\subsection{Morphology and interaction with ISM}
\label{sec:morphology}

The radio morphology of the two lobes is quite different. It is likely
that their asymmetry is related to a slightly different ambient
medium. In particular, the interaction between the relativistic jet
and the ISM can lead to the formation of lobes of various
morphologies.

According to computer simulations presented by Bicknell et
al. (\cite{bic02}), the extended, low brightness morphology of the
western lobe could be produced by a jet interacting with many ISM
clouds. This interaction is not so strong as to destroy the jet but
would cause: (1) the presence of a hot spot fainter than usual and (2)
the spread of a large fraction of relativistic plasma carried by the
jet and therefore the formation of a diffuse, faint lobe.

In the same scenario, the more compact and regular morphology of the
eastern lobe suggests that it is undergoing less interaction with the
ISM. The deep hole close to the hot spot becomes an even more
interesting feature. Could it be due to foreground absorption or is it
truly a large region in the lobe devoid of any radio emitting plasma?
The steep spectral index of 1.7 $\pm$ 0.04 found from the spectral
index image (Fig. \ref{fig7}), argues against foreground absorption by
thermal electrons which would produce an inverted spectrum.  This
leaves us with a remarkable hole in the lobe.  It would be interesting
to search for the presence of dense molecular gas that might be
impenetrable by the radio plasma.

\subsection{Evolution of CSOs}

Despite uncertainties and some discrepancy between kinematic and
spectral age for 4C\,31.04, our results clearly indicate that it is
indeed a very young source. Most of the known Compact Symmetric
Objects have been selected from flux limited samples, so that they are
usually quite bright and they will likely evolve into FR\,II radio
galaxies. It is interesting to ask what will be the fate of the weaker
4C\,31.04, and try to learn more on the relation between `child' and
`adult' stages of radio sources.

One of the problems with the youth scenario is given by the relative
numbers of small sources (CSOs, CSS) and large classical double radio
galaxies. This has been explained by models (e.g. Fanti et
al. \cite{fan95}) in which the total power $P$ of the source decreases
as a power law function of linear size $LS$: $P \propto (LS)^{-h}$. As
a consequence of variations of external density and pressure, the
index $h$ is not very strictly constrained and it may actually change
on different scales and for different sources.

If we roughly use $P \propto (LS)^{-0.5}$ as a law for radio power
evolution (O'Dea \cite{ode98}), we find that 4C\,31.04 will drop below
the power of $10^{24.5}$ W/Hz as soon as it will be $\sim$40 times
larger than its actual size. In other words, by the time 4C\,31.04
will be 5 kpc large, its power will be similar to that of edge
darkened FR\,I radio galaxies.  Even if this threshold is actually not
so sharp and the numbers vary, it is interesting to speculate that
4C\,31.04 may have an FR\,I morphology in the future. In fact, based
on pc-scale jet velocities and properties, \cite{gio01} argued that
there are no significant differences between FR\,I and FR\,II
radiogalaxies on pc-scales. It is then interesting to note that a
possible future FR\,I radio galaxy now has the same edge brightened
look of CSOs evolving into FR\,IIs. Thus, even in a completely
different evolutionary stage, there would seem to be no difference on
the small scale between FR\,Is and FR\,IIs. The discovery of other
CSOs among samples of low-power sources is needed to better understand
the evolution of low-power radio galaxies.

\subsection{Lack of Polarization}

Synchrotron emission is intrinsically polarized up to 70\%, depending
on the amount of order in the source magnetic fields (Burn
\cite{bur66}).  Why is there so little polarized emission from
4C\,31.04 and other CSOs (Peck \& Taylor \cite{pec00a})? From the
H\kern0.1em{\sc i}\ observations we know that 4C\,31.04 is viewed
through a moderate density of atomic gas.  If there is ionized gas
present as well along with a magnetic field, then the radiation from
4C\,31.04 that we observe could have been depolarized by a change in
polarization angle within each 8 MHz IF, or by a gradient in the
foreground Faraday screen.  For bandwidth depolarization to be
effective at 5 GHz requires RMs of 10$^5$ rad m$^{-2}$, while
beam-width depolarization requires gradients in the RM of 300 rad
m$^{-2}$ mas$^{-1}$ or more.

\section{Conclusions}

From the comparison of the two VLBI images at 5 GHz, a small but
significant expansion appears to be present in the Compact Symmetric
Object 4C\,31.04. From the derived expansion rate and the current size
of the source we calculate an age of 550 years. A somehow larger value
of the age (3000 -- 5000 yrs) is derived from spectral index
analysis. Despite a significant difference, both results agree 
that the source is very young, as already demonstrated for other
CSOs in the literature. However, the relatively high hot spots
velocity and its low power make of 4C\,31.04 a peculiar case
among CSOs.

The discrepancy and uncertainty in the age estimates, the uncommon
presence of a deep `hole' in the East lobe, the detection and the
variability of the compact core suggest that it may be worth observing
this source again in the future. New VLBI observations will be of
interest to look for free--free absorption at low frequency, to
confirm the expansion, and to look for spectral aging at high
frequency.

\begin{acknowledgements}
We thank the referee, Dr. M. S. Brotherton, for useful comments which
improved the clarity of the paper.  This research has made use of the
NASA/IPAC Extragalactic Database (NED) which is operated by the Jet
Propulsion Laboratory, Caltech, under contract with NASA. This
research has also made use of NASA's Astrophysics Data System Abstract
Service.  GBT thanks the CNR for hospitality during his visit when
much of this work was accomplished. This work was partly supported by
the Italian Ministry for University and Research (MIUR) under grant
COFIN 2001-02-8773.
\end{acknowledgements}

\end{document}